\documentclass[aps,prb,twocolumn,showpacs,floatfix]{revtex4}
\usepackage{graphicx}
\usepackage{textcomp}
\begin{document}

\title{Structure and vibrational spectra of carbon clusters in SiC}

\author{Alexander Mattausch}
\email{Alexander.Mattausch@physik.uni-erlangen.de}

\author{Michel Bockstedte}

\author{Oleg Pankratov}

\affiliation{Lst. f\"ur theoretische Festk\"orperphysik, Universit\"at
  Erlangen-N\"urnberg, Staudtstr. 7, 91058 Erlangen, Germany}
\date{\today}

\begin{abstract}
  The electronic, structural and vibrational properties of small
  carbon interstitial and antisite clusters are investigated by
  \emph{ab initio} methods in 3C and 4H-SiC. The defects possess
  sizable dissociation energies and may be formed via condensation of
  carbon interstitials, e.g.\ generated in the course of ion
  implantation. All considered defect complexes possess localized
  vibrational modes (LVM's) well above the SiC bulk phonon spectrum.
  In particular, the compact antisite clusters exhibit high-frequency
  LVM's up to 250\,meV. The isotope shifts resulting from a $^{13}$C
  enrichment are analyzed. In the light of these results, the
  photoluminescence centers D$_{\text{II}}$ and P$-$U are discussed.
  The dicarbon antisite is identified as a plausible key ingredient of the
  D$_{\text{II}}$-center, whereas the carbon split-interstitial is a
  likely origin of the P$-$T centers. The comparison of the calculated
  and observed high-frequency modes suggests that the U-center is also
  a carbon-antisite based defect.
\end{abstract}


\bibliographystyle{apsrev}

\maketitle

\section{Introduction}
\label{sec:introduction}

Silicon carbide is a wide-band-gap semiconductor material especially
suitable for high power, high frequency and high temperature
applications. A common technique for the incorporation of dopant atoms
into SiC is ion implantation. The post-implantation annealing is
needed to reduce the damage. At the same time, the thermal treatment
creates new stable defect centers which influence the material
properties. The persistent centers that tend to grow even at annealing
temperatures above 1000$^{\circ}$C are e.g.\ the photoluminescence
(PL) centers D$_{\text{I}}$ (Ref.~\onlinecite{Pa72}) and
D$_{\text{II}}$ (Ref.~\onlinecite{Pa73}).

Most likely, the formation of these centers is related to the
aggregation of the intrinsic defects. For example, vacancy or antisite
clusters have been discussed\cite{Pa72,Eb02,Ga03} in conjunction with
the D$_{\text{I}}$-center. The D$_{\text{II}}$-center, with its
rich vibrational spectrum located above the SiC phonon spectrum, is
considered as a carbon-related defect.\cite{Pa73,Ma03,Ga03} Up to now,
the complexity of the vibrational spectrum prevents an unambiguous
identification of the microscopic structure. However, the theoretical
investigations of the vibrational spectrum of small carbon clusters
indicates that the dicarbon antisite (a carbon pair at the silicon
site) is an important structural element of this defect.\cite{Ma03}
Besides the D$_{\text{II}}$-center, further carbon-related defects
have been identified recently.\cite{Ev02,St03_priv} The
photoluminescence centers P$-$U appear in irradiated material and
vanish between 1000$^{\circ}$C$-$1300$^{\circ}$C.\cite{Ev02} Based on
the analysis of $^{13}$C isotope shifts of the LVM frequencies, a
carbon split-interstitial model was suggested for the P$-$T
centers.\cite{Ev02} This proposal was recently questioned\cite{Ga03a}
and an alternative model was proposed. Generally, although the exact
microscopic models for these defects remain to be established, it is
clear that they all are due to the aggregation of carbon interstitials
in smaller or larger clusters, which constitute a fundamentally
important class of defects in SiC.

The finding, that the carbon interstitials aggregate, can help to
resolve a controversy regarding the formation of the persistent defect
centers. Namely, the predicted high mobility of the carbon
split-interstitials\cite{Ma01,Bo03b} should hinder the formation of
D$_{\text{II}}$ centers at high temperatures\cite{Pa73,Fr87} and
opposes the thermal stability of the P$-$T centers.\cite{Ev02} Due to
the high mobility, the carbon interstitials, which are a key
ingredient of these centers,\cite{Pa73,Ma03,Ev02} should have vanished
at elevated temperatures. Nevertheless, both features are observed
experimentally. On the other hand, the recent tentative
assignment\cite{Pe02,Bo03a,Bo03,Ga03a} of the electron paramagnetic
resonance (EPR) centers T5 in 3C-SiC (Ref.~\onlinecite{It97a}), and
EI1 and EI3 (Ref.~\onlinecite{So01a}) in 4H-SiC to carbon
split-interstitials, which is based on the comparison of calculated
and measured hyperfine parameters, supports the prediction of the high
mobility. These centers anneal between 150$^{\circ}$C and
240$^{\circ}$C. A low temperature annealing of the silicon and carbon
interstitials was also suggested for irradiated or implanted material
by our theoretical analysis of competing annealing
mechanisms.\cite{Bo03,Bo04} The interstitial migration and the
vacancy-interstitial recombination are activated at much lower
temperatures than other mechanisms. We recently proposed\cite{Bo04}
that small carbon aggregates may provide the missing link between the
carbon interstitials and the persistent carbon-related defects. Such
aggregates may trap interstitials at lower temperatures and re-emit
them again at higher temperatures, or even represent persistent
photoluminescence centers themselves. Also in the transient enhanced
dopant diffusion (TED) the carbon aggregates may serve as a source of
carbon interstitials similarly to the so-called \{311\}-defects in
silicon,\cite{Ea94} which emit silicon interstitials and thereby
facilitate the TED.

A thorough investigation of small and larger carbon aggregates should
provide an important information concerning their role in the
annealing of implantation damage. Theoretical prediction of their
vibrational spectra could also help to identify the microscopic
structure of the PL-centers. Yet, the assignment of a particular
defect structure to PL centers from the phonon replicas alone is
difficult. As an additional information, the thermal stability and
growth/annealing mechanisms of the clusters may be investigated by
analyzing the dissociation energy of the defect complex and its
preceding configurations. Sofar, theoretical investigations focused on
di-interstitials and small interstitial clusters at
antisites.\cite{Ma03,Ga03a} A detailed discussion of the vibrational
signature of larger clusters is still missing.

In the present article we investigate the structure, dissociation
energies and vibrational properties of carbon interstitial and
antisite aggregates and of larger clusters involving up to four carbon
interstitials. We employ an \emph{ab initio} method within the
framework of density functional theory. The isotope shifts of defects
with a distinct and clear splitting pattern are presented.  Although
theoretically all these defects can acquire different charge states,
we focus on the neutral state as the most relevant for PL experiments.
The PL due to bound excitons at a charged defect is unlikely to be
seen as more non-radiative recombination channels are open in this
case.\cite{St75} We have found high-frequency LVMs with up to 250\,meV
for small and compact defect clusters at antisites and LVMs up to
200\,meV for interstitial clusters. As the cluster size increases, the
frequency of the highest mode drops. In general, we find that the bond
length between the involved carbon atoms, which is a measure of the
bond strength, plays a crucial role. The analysis of further
antisite-based clusters provides additional support for the previously
suggested model of the dicarbon antisite as a core structure of the
D$_{\text{II}}$ center.\cite{Ma03} The vibrational signature and the
isotope shifts of the split-interstitial in the tilted configuration
favor this defect as a candidate for the P$-$T centers, although an
assignment to the dicarbon antisite cannot be completely rejected. The
existence of LVMs up to 250\,meV for compact antisite clusters shows
that the original suggestion of a completely carbon-based $U$-center
is possible. A comparison with theoretical results for
diamond\cite{Go01} reveals that interstitial defects with similar
behavior as in SiC are also found in this structurally very similar
material.

The article is organized as follows: Section~\ref{sec:method} is
denoted to a description of the method. In
Secs.~\ref{sec:interst-clust} and~\ref{sec:antisite-clusters} we
present our results for interstitial and antisite clusters in detail
and illustrate the rich variety of possible defect structures. In
Sec.~\ref{sec:discussion} the results will be discussed in the light
of experimental and theoretical data. A summary concludes the article.

\section{Method}
\label{sec:method}

Our first principles calculations have been carried out using the
plane wave pseudopotential package \textsf{FHI96SPIN}\cite{Bo97}
within the density functional theory (DFT) formalism.\cite{Ho64,Ko65}
The details of the method can be found in Ref.~\onlinecite{Bo03b}.
For the expansion of the Kohn-Sham wave functions a basis set of
30\,Ryd cut-off energy is used. The exchange-correlation potential
taken in the local density approximation (LSDA).\cite{Pe81} For the
calculation of the defect energetics large supercells with 216 lattice
sites have been employed for 3C-SiC and with 128 lattice sites for
4H-SiC. The Brillouin zone has been sampled at the $\Gamma$ point in
3C-SiC, in 4H-SiC a $2\times2\times2$ Monkhorst-Pack mesh\cite{Mo76}
has been used. All atoms in the cell have been allowed to relax.

The abundance of a defect under equilibrium conditions is given by its
formation energy. However, as the radiative production of the defects
is not an equilibrium process, the relevance of the formation energy
is rather limited. For the formation and the stability of the carbon
clusters in damaged material the dissociation energy is the more
important quantity. It is defined as an energy needed to remove a
single atom from the defect. This quantity describes the thermal
stability of a defect. In practice, for the carbon clusters, the
dissociation energy is the energy difference between the total energy
of the cluster and the total energy of the cluster with one carbon
atom less plus the energy of an infinitely remote carbon interstitial:
\begin{equation}
  \label{eq:Ed}
  E_{\text{D}} = E_{\text{tot}}(\text{C}_{n-1}) +
  E_{\text{tot}}(\text{C}_{\text{sp}}) - E_{\text{tot}}(\text{C}_{n}) 
\end{equation}

Here C$_{\text{sp}}$ denotes the $\langle 100 \rangle$ carbon
split-interstitial, which is the energetically lowest carbon
interstitial. Since the most stable configuration of the isolated
carbon defect is used as the reference energy, the Eq.~(\ref{eq:Ed})
gives the lower bound of the dissociation barrier. We note that the
tilted $\langle 100 \rangle$ split-interstitial configuration has been
chosen, which is about 0.5\,eV lower in energy than the symmetric
$\langle 100 \rangle$-geometry.\cite{Bo03b} The corresponding values
of the formation energies are 6.7\,eV in 3C-SiC and 7.3\,eV in 4H-SiC.
For all dissociation energies the neutral charge state has been
assumed, as the carbon clusters are typically neutral in a typical
range of the Fermi level (cf.\ Secs.~\ref{sec:interst-clust} and
\ref{sec:antisite-clusters}).

The calculation of the vibrational modes has been performed in large
supercells employing the defect molecule approximation. In this
approximation the atoms belonging to the defect as well as the nearest
neighbors are systematically displaced by about 0.15$-$0.7\% of the
bond length and the dynamical matrix is calculated from the energy
variation. The forces have been converged to a relative accuracy of
$5\times 10^{-5}$. A detailed outline of the technique is given in
Ref.~\onlinecite{Ma03}. For the calculation of the dynamical matrix in
4H-SiC the Brillouin zone has been sampled by the $\Gamma$ point. It
has been verified that the accuracy of the LVM calculation is not
affected by this change. As we have shown earlier,\cite{Ma03} the main
effect on the accuracy results from the uncertainty in the lattice
constant (the lattice constant of the supercell including the defect
differs slightly from that of the ideal lattice) and from the defect
molecule approximation (which affects the energetically lowest LVMs
close to the bulk spectrum). Due to the former the LVMs may change up
to 10\,meV when varying the lattice constant from the experimental
value to the (smaller) theoretical LDA value. The latter results in a
lowering of the energetically lowest LVMs. In some cases, referred to
in the text, a 64 sites cell has been used to calculate the LVMs. In
these cases all atoms of the cell were included into the dynamical
matrix.

The most important feature of PL centers is the zero phonon line (ZPL)
that originates from a direct recombination of the bound exciton
without involving the LVM's. Unfortunately, an analysis of the
recombination process is beyond the capabilities of the static
DFT-based methods. We therefore do not consider the recombination by
itself, but focus on the vibrational modes as fingerprints of the
particular carbon aggregates.

\section{Interstitial clusters}
\label{sec:interst-clust}

Interstitial clusters are the simplest carbon aggregates. The original
lattice structure is not altered, no antisites or vacancies are needed
for the cluster formation. The only prerequisite is an availability of
carbon interstitials. Due to a large formation energy of more than
6.5\,eV the equilibrium concentration of these defects is well below
the detectable limit. Hence the formation of the carbon clusters in a
noteworthy concentration is possible only in irradiated or implanted
material.

\begin{table*}
  \caption{\label{tab:iso-csp} LVMs (in meV) and isotope shifts of the
    carbon split-interstitial dumbbell C$_{\text{sp}\langle 100
    \rangle}$ in the tilted configuration in 3C and 4H-SiC. The values
    for 3C-SiC in parentheses are obtained using the defect-molecule
    approximation. As the tilted configuration of C$_{\text{sp}\langle 
      100 \rangle}$ is asymmetric, the frequencies with only one
    substituted dumbbell atom depend on the orientation. The notation
    $^{12}$C$-^{13}$C  denotes the dumbbell configuration where the
    $^{13}$C-atom is sp$^{3}$ hybridized (cf.\ text).}
  \begin{ruledtabular}
    \begin{tabular}{ccccccccccccccc}
                        & \multicolumn{4}{c}{3C} &
      \multicolumn{4}{c}{4H, cubic} & \multicolumn{6}{c}{4H, hexagonal}\\
      \cline{2-5} \cline{6-9}       \cline{10-15}
      Isotopes          & LVM   & Ratio & LVM   & Ratio & LVM   &
      Ratio & LVM   & Ratio & LVM   & Ratio & LVM   & Ratio & LVM &
      Ratio \\
      \hline                                                                            
      $^{12}$C$-^{12}$C & 188.7 (187.9) &       & 120.4 (110.9) &       & 186.6 &  
      & 111.3 &       & 192.9  &        & 124.1 &       & 122.5 & \\
      $^{13}$C$-^{12}$C & 185.7 (185.0) & 1.016 & 120.2 (109.7) &
      1.002 (1.011) & 183.6 & 1.016 & 110.4 & 1.008 & 189.2 & 1.020 &
      123.9 & 1.002 & 119.1 & 1.028 \\
      $^{12}$C$-^{13}$C & 184.7 (183.9) & 1.022 & 120.0 (109.2) &
      1.003 (1.016) & 182.6 & 1.022 & 109.0 & 1.021 & 189.3 & 1.019 &
      122.9 & 1.011 & 119.8 & 1.022 \\
      $^{13}$C$-^{13}$C & 181.6 (180.8) & 1.039 & 119.8 (108.0) &
      1.005 (1.027) & 179.5 & 1.039 & 108.1 & 1.029 & 185.5 & 1.040 &
      120.3 & 1.032 & 118.8 & 1.031 \\
    \end{tabular}
  \end{ruledtabular}

\end{table*}

\subsection{Carbon split-interstitial}
\label{sec:carb-split-interst}

The carbon split-interstitial C$_{\text{sp}}$ (two carbon atoms
sharing a single carbon lattice site) is the energetically lowest of
the carbon interstitials. It possesses the charge states $2^{+}$ to
$2^{-}$. As discussed in Ref.~\onlinecite{Bo03b}, the carbon
split-interstitial can be found in two different configurations. In
the positive charge states, the dumbbell is oriented in $\langle 100
\rangle$ direction and the defect has the $D_{\text{2d}}$ symmetry,
which may be lowered to $D_{2}$ due to the Jahn-Teller effect. In the
symmetric configuration, two dangling $p$-orbitals are oriented in
$\langle 110 \rangle$ direction. In the neutral and negative charge
states, the defect gains about 0.5\,eV by adopting a tilted structure
compared to the metastable symmetric configuration. The $\langle 100
\rangle$ dumbbell tilts by about 30$^{\circ}$, so that one of the two
carbon atoms acquires a sp$^{3}$ bonding configuration with the three
silicon neighbors. In 4H-SiC two different configurations of the
split-interstitial should be considered. The alternating cubic and
hexagonal planes in the 4H stacking sequence allows for two
inequivalent sites---cubic and hexagonal---available to the defect. At
a cubic site, the next-nearest neighbor configuration is the same as
in the zinc-blende structure, whereas at a hexagonal site it is of a
wurtzite-type. The tilt angles of the split-interstitial dumbbell in
3C-SiC and in 4H-SiC at the cubic site are similar. At the hexagonal
site the tilt is also visible, but less pronounced. At both sites in
4H-SiC, the energy gain due to the tilt is only about 0.2\,eV, which
is less than in 3C-SiC. The driving force for the tilting is the
stretching of the tetrahedron along the $c$-axis already in the
undistorted crystal, so that in 4H-SiC the undistorted configuration
is closer to the tilted configuration than in 3C-SiC.

The tilted configuration of the split-interstitial should be visible
via specific phonon replica in photoluminescence experiments. The
calculated LVMs and the isotope shifts are listed in
Tab.~\ref{tab:iso-csp}. The calculation has been performed using full
64 sites supercells with a special $2\times2\times2$
$\mathbf{k}$-point sampling in 3C-SiC, whereas the defect molecule
approximation has been employed for 4H-SiC.  For the 3C polytype we
found two vibrational modes above the SiC bulk phonon spectrum at
188.7\,meV and 120.4\,meV. A similar result is obtained for the
split-interstitial at the cubic site (C$_{\text{sp,k}}$). In this case
we also find two LVMs at 186.6\,meV and 111.3\,meV. Since in 3C-SiC
the defect molecule approximation underestimates the frequency of the
lower LVM by 10\,meV (cf.  Tab.~\ref{tab:iso-csp}), we expect a
similar shift of the modes in 4H-SiC. The higher LVM can be described
as an isolated stretching vibration of the dumbbell, whereas the lower
LVM is a concerted motion of the complete dumbbell against the
elongated edge of the distorted tetrahedron. At the hexagonal site
(C$_{\text{sp,h}}$), the vibrational pattern is slightly different. We
obtain a stretching vibration with an energy of 192.9\,meV, whereas
there are two modes with energies 124.1\,meV and 122.9\,meV. They are
also affected by the defect molecule approximation, thus they are
likely to rise in energy with the inclusion of additional surrounding
shells in the calculation.\cite{Ma03} The nearly degenerate modes
describe vibrations of one of the carbon atoms against its two silicon
neighbors. The lifting of the degeneracy of these modes reflects the
structural distortion, which is less pronounced at the hexagonal site
than at the cubic site. These results may as well be transfered to
6H-SiC, which crystallizes with a hexagonal-cubic-cubic stacking. The
cubic plane's environment is 3C-like, whereas for the hexagonal plane
the same 2H-like structure as for the hexagonal plane in 4H-SiC is
present. Using the symmetric configuration of the split-interstitial,
Gali \emph{et al.}\cite{Ga03a} obtained only a single LVM at 197\,meV
in 3C-SiC and at 183\,meV for the hexagonal interstitial in 4H-SiC.

The calculated LVM's (including isotope shifts due to a substitution
of one or both atoms with $^{13}$C) are summarized in
Tab.~\ref{tab:iso-csp}. It is seen that the high-frequency LVM splits
into four lines, with the two medium lines separated by only 1\,meV.
The frequency ratios for the shifted modes are close to the square
root of the reduced mass ratios, which is 1.020 for the
$^{12}$C$-^{13}$C vibration and 1.041 for the $^{13}$C$-^{13}$C
vibration. For the low-frequency LVM, the two $^{12}$C$-^{13}$C modes
shift asymmetrically with the substitution of one carbon atom. The
asymmetry is due to the tilted geometry of the defect. We observe in
our results for 3C-SiC that the frequency ratio for this mode is
affected by the defect molecule approximation. When the full supercell
is included into the calculation, the frequency shifts reduce to less
than 1\,meV, resulting in a much lower frequency ratio.  We expect a
similar effect for the cubic site in 4H-SiC.

\begin{figure*}
  \includegraphics[width=\linewidth]{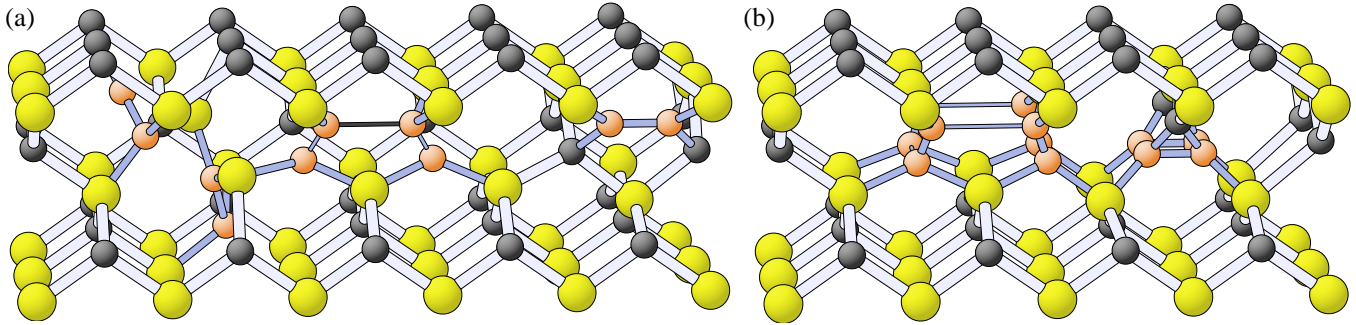}

\caption{\label{fig:3C-int} Carbon-interstitial clusters in
  3C-SiC. (a) Carbon di-interstitials
  (C$_{\text{sp}}$)$_{2,\text{lin}}$, (C$_{\text{sp}}$)$_{2}$, and
  (C$_{2}$)$_{\text{Hex}}$. (b) Carbon tetra-interstitials
  (C$_{\text{sp}}$)$_{4}$ and ((C$_{2}$)$_{\text{Hex}}$)$_{2}$.}
\end{figure*}

\begin{table}
  \caption{  \label{tab:cluster-bind} Dissociation energy of the
    neutral carbon clusters. The given value is the energy needed to
    remove a single carbon atom. For the cluster type
    (C$_{3}$)$_{\text{Si}}$, the labels $I1$ and $I2$ refer to the
    metastable intermediate configurations, whereas the most stable
    triangular geometry is labeled
    (C$_{3,\text{tri}}$)$_{\text{Si}}$. The two values for
    (C$_{\text{sp}}$)$_{4,kkhh}$ in 4H-SiC denoted with \emph{f. hex.}
    and \emph{f. cub.} are the energies needed to remove an atom from
    the hexagonal or cubic site, respectively.}
  \begin{ruledtabular}
    \begin{tabular}{cccccc}
      cluster type & \multicolumn{5}{c}{Dissociation energy (eV)} \\ 
      \cline{2-6}
      3C-SiC &&&&\\
      (C$_{\text{I}}$)$_{2}$ & (C$_{\text{sp}}$)$_{2,\text{lin}}$ &
      1.9 && (C$_{\text{sp}}$)$_{2}$ & 2.8\\
      & (C$_{2}$)$_{\text{Hex}}$ & 4.8 &&&\\
      (C$_{\text{I}}$)$_{3}$ &
      (C$_{2}$)$_{\text{Hex}}-$C$_{\text{sp}}$ & 0.4 &&
      (C$_{\text{sp}}$)$_{3}$ & 3.0 \\
      (C$_{\text{I}}$)$_{4}$ & ((C$_{2}$)$_{\text{Hex}}$)$_{2}$ & 4.0 &&
      (C$_{\text{sp}}$)$_{4}$ & 5.7 \\ 
      (C$_{2}$)$_{\text{Si}}$ &          & 4.1      &&& \\
      (C$_{3}$)$_{\text{Si}}$ & $I1$ & 1.4 && $I2 $ & 3.6\\
      & (C$_{3,\text{tri}}$)$_{\text{Si}}$ & 4.8 &&&\\
      (C$_{4}$)$_{\text{Si}}$ &          & 2.8      &&& \\
      ((C$_{2}$)$_{\text{Si}}$)$_{2}$ &  & 5.9      &&& \\
      4H-SiC&&&&\\
      (C$_{\text{I}}$)$_{2}$  & (C$_{\text{sp}}$)$_{2,\text{kk}}$ &
      2.5 && (C$_{\text{sp}}$)$_{2,\text{hh}}$ & 5.5 \\
      & (C$_{\text{sp}}$)$_{2,\text{kk,lin}}$  & 4.6 &&
      (C$_{\text{sp}}$)$_{2,\text{hh,lin}}$ & 4.6 \\
      & (C$_{\text{sp}}$)$_{2,\text{hk,cub}}$ & 4.2 &&
      (C$_{\text{sp}}$)$_{2,\text{kh,hex}}$ & 5.4 \\

      & (C$_{\text{2}}$)$_{\text{Hex,k}}$ & 5.1
      &&(C$_{\text{2}}$)$_{\text{Hex,k}}$ & 5.1\\
      (C$_{\text{I}}$)$_{3}$ & (C$_{\text{sp}}$)$_{3,\text{kkh}}$ &
      2.9 && (C$_{\text{sp}}$)$_{3,\text{khh}}$ & 1.3 \\
      (C$_{\text{I}}$)$_{4}$ & (C$_{\text{sp}}$)$_{4, \text{kkhh}}$  &
      5.3 &(\emph{f. hex.}) &\\
      && 4.8 &(\emph{f. cub.}) &&  \\
      (C$_{2}$)$_{\text{Si}}$ & (C$_{2}$)$_{\text{Si,h}}$ & 3.6 &&
      (C$_{2}$)$_{\text{Si,k}}$ & 3.5 \\
      (C$_{3}$)$_{\text{Si}}$ & $I1$ & 1.9 &&
      (C$_{3,\text{tri}}$)$_{\text{Si,k}}$ & 5.8 \\
      (C$_{4}$)$_{\text{Si}}$ &      & 2.9 &&     & \\
      ((C$_{2}$)$_{\text{Si}}$)$_{2}$ &
      ((C$_{2}$)$_{\text{Si}}$)$_{2,\text{kh}}$ & 6.7  && \\
    \end{tabular}
  \end{ruledtabular}
\end{table}

\subsection{Carbon di-interstitials}
\label{sec:carb-di-interst}

The unsaturated $p$-orbitals of the interstitials enable the formation
of stable di-interstitial complexes. As described in detail in
Ref.~\onlinecite{Bo04}, different forms of the di-interstitial are
possible depending on the orientation of the neighboring interstitial
sites. If the two interstitials are not contained within the
\{110\}-plane, a complex reconstruction occurs as depicted in
Fig.~\ref{fig:3C-int}a left. Since this defect consists of two tilted
split-interstitials and a strongly displaced neighboring silicon atom,
which results in a nearly linear geometry of carbon interstitials, we
have labeled this di-interstitial (C$_{\text{sp}}$)$_{2,\text{lin}}$
(for consistency with the di-interstitials in 4H-SiC, we have dropped
the earlier notation (C$_{\text{sp}}$)$_{2,\text{tilted}}$ of
Ref.~\onlinecite{Bo04}). With 1.9\,eV this defect possesses the lowest
dissociation energy of all the di-interstitials in 3C-SiC discussed
here (cf.  Tab.~\ref{tab:cluster-bind}). The di-interstitial
(C$_{\text{sp}}$)$_{2}$ (Fig.~\ref{fig:3C-int}a center), which is
comprised of two neighboring carbon split-interstitials that are
inclined towards each other, is more stable with a dissociation energy
of 2.8\,eV. The most stable reconstruction is (C$_{2}$)$_{\text{Hex}}$
(Fig.~\ref{fig:3C-int}a right) with a dissociation energy of 4.8\,eV,
where the di-interstitial is contained in a hexagonal ring. As the
lattice geometry is only slightly distorted by this di-interstitial,
the term (C$_{2}$)$_{\text{Hex}}$ denotes that the di-interstitial is
enclosed by a hexagonal ring. For this defect Gali \emph{et al.}
obtained a higher dissociation energy of 5.3\,eV,\cite{Ga03a} which,
however, was calculated with a higher reference energy for the carbon
interstitial.

The electronic structure of these defects has been discussed in
Ref.~\onlinecite{Bo04}. For a Fermi level position around the mid-gap
all di-interstitials are neutral. The di-interstitial
(C$_{\text{sp}}$)$_{2,\text{lin}}$ is neutral for
$\mu_{\text{F}}>0.8$\,eV, (C$_{\text{sp}}$)$_{2}$ for
$\mu_{\text{F}}>1.0$\,eV and (C$_{2}$)$_{\text{Hex}}$ for all Fermi
levels. The upper bound of the neutrality range could not be obtained,
since the defect levels are resonant with the (artificially low) LDA
conduction band.

\begin{figure*}
  \includegraphics[width=\linewidth]{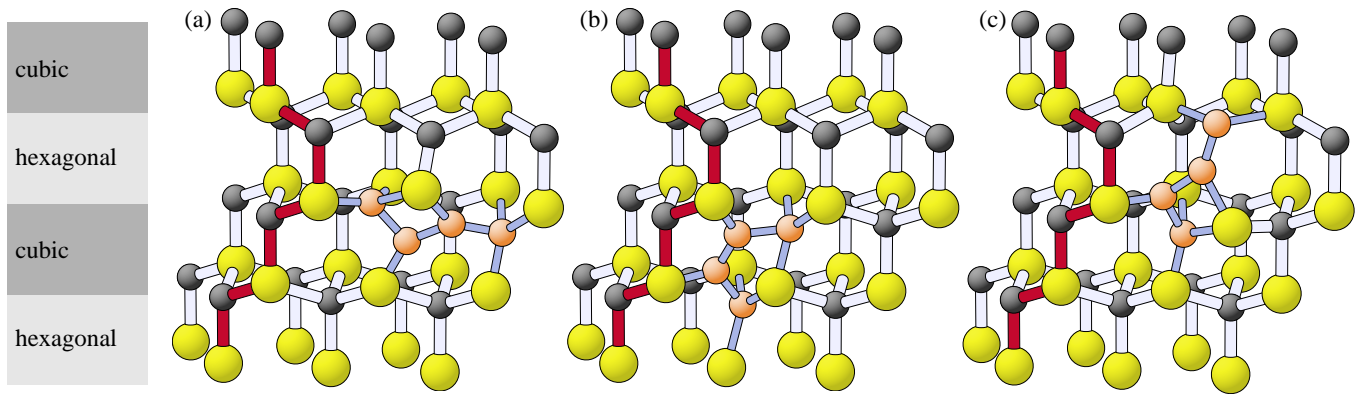}
  \caption{\label{fig:diint-4H} Carbon di-interstitials in 4H-SiC. (a)
    linear cubic-cubic di-interstitial
    (C$_{\text{sp}}$)$_{2,\text{kk,lin}}$ (b) hexagonal-cubic
    di-interstitial in the cubic plane
    (C$_{\text{sp}}$)$_{2,\text{hk,cub}}$ (c) cubic-hexagonal
    di-interstitial in the hexagonal plane
    (C$_{\text{sp}}$)$_{2,\text{kh,hex}}$ The stacking order of the
    crystal is highlighted, the cubic and hexagonal planes are marked
    at the left.}
\end{figure*}

In 4H-SiC a larger variety of di-interstitials is available. In
addition to the various directions of the interstitial pair, the
inequivalency of cubic and hexagonal sites has to be considered.  In
the stacking sequence of 4H-SiC, the cubic 3C-like and the hexagonal
2H-like planes alternate (cf. Fig.~\ref{fig:diint-4H}). Hence a
di-interstitial can comprise two cubic or two hexagonal sites in two
different orientations. Figure~\ref{fig:diint-4H}a shows a
di-interstitial on two cubic sites which is comparable to
(C$_{\text{sp}}$)$_{2,\text{lin}}$ in 3C-SiC, except that the two
split-interstitials show a much stronger relaxation towards each
other. (Note that the $c$-axis in 3C-SiC is oriented along the
$\langle 111 \rangle$ direction.) In this case the defect is totally
contained within a cubic or hexagonal plane.  Alternatively, the pair
can reside on one cubic and one hexagonal site. Again, two different
configurations are possible: the center of the di-interstitial may be
contained in a 3C-like cubic plane (Fig.~\ref{fig:diint-4H}b) or in a
2H-like hexagonal plane (Fig.~\ref{fig:diint-4H}c). Including
di-interstitials within a hexagonal ring ((C$_{2}$)$_{\text{Hex}}$),
which may reside in 4H-SiC within a cubic or a hexagonal plane, eight
different di-interstitial configurations can be counted in total,
instead of only three in 3C-SiC. They possess different dissociation
energies (cf.  Tab.~\ref{tab:cluster-bind}). The least stable
structure is (C$_{\text{sp}}$)$_{2,\text{kk}}$, two
split-interstitials at neighboring cubic sites, whose geometry and
dissociation energy of 2.5\,eV are comparable to its 3C-SiC
counterpart. More stable are the cubic-hexagonal di-interstitial
(C$_{\text{sp}}$)$_{2,\text{kh,hex}}$ (cf. Fig.~\ref{fig:diint-4H}c)
in the hexagonal bi-layer with a dissociation energy of 4.2\,eV and
the linear configurations (C$_{\text{sp}}$)$_{2,\text{kk,lin}}$ and
(C$_{\text{sp}}$)$_{2,\text{hh,lin}}$ with dissociation energies of
4.6\,eV. As in 3C-SiC, the notation C$_{\text{sp}}$ is used where the
original carbon lattice atoms are strongly displaced and share their
site with a carbon interstitial. The most stable configurations of the
di-interstitial are those that are contained within a hexagonal ring.
This is not only the case for both (C$_{2}$)$_{\text{Hex,k}}$ and
(C$_{2}$)$_{\text{Hex,h}}$ (the hexagonal di-interstitials located in
cubic and hexagonal planes), but also for
(C$_{\text{sp}}$)$_{2,\text{kh,cub}}$ (Fig.~\ref{fig:diint-4H}b) and
(C$_{\text{sp}}$)$_{2,\text{hh}}$. For the latter two configurations
the di-interstitial is rotated so that it enters the neighboring
hexagonal ring, which results in a structure similar to
(C$_{2}$)$_{\text{Hex}}$. The term C$_{\text{sp}}$ in these
di-interstitials reflects more the starting geometry than the final
configuration, which is depicted in Fig.~\ref{fig:diint-4H}b, since
the two carbon interstitials relax strongly into the nearby hexagonal
ring. With dissociation energies of 5.1\,eV [(C$_{2}$)$_{\text{Hex}}$
within the cubic and hexagonal plane], 5.4\,eV
[(C$_{\text{sp}}$)$_{2,\text{kh}}$] and 5.5\,eV
[(C$_{\text{sp}}$)$_{2,\text{hh}}$] these di-interstitials are the
most stable di-interstitial structures in 4H-SiC. All defects are
electrically neutral for the Fermi-level position between
$\mu_{\text{F}} > 0.5$\,eV and $\mu_{\text{F}} < 2.5$\,eV.

\begin{table*}
    \caption{\label{tab:int-clust-lvm} LVMs of neutral interstitial
      clusters in 3C and 4H-SiC in meV. The values marked by an
    asterisk are slightly below the bulk phonon limit, which may be a
    result of the defect-molecule approximation. The edge of the 
    bulk phonon spectrum obtained from \emph{ab initio} calculations
    neglecting the macroscopic crystal polarization is 115.5\,meV for 3C-SiC
    and 118.6\,meV for 4H-SiC.}
  \begin{ruledtabular}
    \begin{tabular}[t]{cccccccccc}
      &\multicolumn{5}{c}{3C}&\multicolumn{4}{c}{4H}\\
      \cline{2-6}\cline{7-10}
      LVM & (C$_{\text{sp}}$)$_{2}$ & (C$_{2}$)$_{\text{Hex}}$ &
      (C$_{\text{sp}}$)$_{3}$ & (C$_{\text{sp}}$)$_{4}$ &
      ((C$_{2}$)$_{\text{Hex}}$)$_{2}$ &
      (C$_{\text{sp}}$)$_{2,\text{kk}}$ &
      (C$_{\text{sp}}$)$_{2,\text{hk,cub}}$ &
      (C$_{\text{sp}}$)$_{2,\text{hh}}$ &
      (C$_{\text{sp}}$)$_{2,\text{kk,lin}}$  \\
      \hline
      1 & 122.5 & 132.5 & 120.7 & 119.7 & 113.2*& 123.7  & 128.0 & 134.9 & 117.1* \\
      2 & 170.3 & 160.4 & 127.5 & 137.2 & 121.0 & 126.3 & 161.3 & 159.9 & 120.2 \\
      3 & 185.1 & 167.3 & 136.6 & 139.5 & 139.0 & 159.3 & 167.1 & 168.1 & 160.0 \\
      4 &       & 184.3 & 161.0 & 139.5 & 142.2 & 164.6 & 189.1 & 191.5 & 167.0 \\
      5 &       &       & 163.2 & 142.4 & 150.2 &       &       &       & 203.2 \\
      6 &       &       &       &       & 160.1 &       &       &       &       \\
      7 &       &       &       &       & 194.7 &       &       &       &       \\
      8 &       &       &       &       & 195.7 &       &       &       &       \\
    \end{tabular}
  \end{ruledtabular}    
\end{table*}

The LVMs of the investigated interstitial clusters are listed in
Tab.~\ref{tab:int-clust-lvm}. The di-interstitial's LVMs lie well
above the SiC bulk phonon spectrum.  With three high-frequency LVMs up
to 185.1\,meV (C$_{\text{sp}}$)$_{2}$ possesses the most simple LVM
spectrum. In 4H-SiC this defect shows four LVMs with the highest mode
at 164.6\,meV, nearly 20\,meV below its counterpart in 3C-SiC. The
reason for this frequency shift is the softening of the defect's bonds
in 4H-SiC. Indeed, the bond length between the split-interstitials is
in 4H-SiC about 0.11\,\AA\ longer than in 3C-SiC. Gali \emph{et
  al.}\cite{Ga03a} obtained for the same defect in 3C-SiC two
high-frequency LVMs at 174\,meV and 165\,meV and four LVMs between
125\,meV and 123\,meV. The discrepancies in the high-frequency LVMs
may be attributed to geometrical variations, whereas for the
low-frequency LVMs this may be caused by the defect molecule
approximation. Yet, the spectrum structure is similar for all three
defects: two relatively close high-frequency LVMs and LVMs close to
the highest SiC bulk phonon mode. Also the the vibrational pattern of
the defects in 3C-SiC and 4H-SiC is similar. The two highest modes are
the stretching vibrations of the two dumbbells (modes 2 and 3 in
3C-SiC and modes 3 and 4 in 4H-SiC in Tab.~\ref{tab:int-clust-lvm}).
The low-energy modes involve vibrations against the nearest neighbor
atoms of the defect. In both polytypes the defect shows complex
isotope shifts with an at least fivefold splitting of the highest LVM
and an at least fourfold splitting of the second highest LVM, which
partially overlap.

The di-interstitials (C$_{2}$)$_{\text{Hex}}$ in 3C-SiC and
(C$_{\text{sp}}$)$_{2,\text{hk,cub}}$ and
(C$_{\text{sp}}$)$_{2,\text{hh}}$ in 4H-SiC reveal their common
hexagonal nature also in their vibrational modes: the maximum
deviation between the modes of the defects is about 7\,meV, which lies
within the accuracy of the method.\cite{Ma03} Due to their structural
analogy, we expect similar results for the di-interstitials
(C$_{2}$)$_{\text{Hex}}$ within the cubic and hexagonal planes.  The
modes of the hexagonal di-interstitials reach up to 191.5\,meV. This
highest mode 4 is a stretching vibration of the two carbon atoms
enclosed in the hexagonal ring, whereas the modes 2 and 3 are
symmetric and antisymmetric stretching vibrations with the two
adjacent carbon atoms. The mode 1 is a rotation of the two outer
dumbbells against each other. The defects in 4H-SiC show the same
vibrational pattern. Comparable results have been found by Gali
\emph{et al.}\cite{Ga03a} with LVMs at 190\,meV, 172\,meV, 166\,meV,
135\,meV, 133\,meV and 117\,meV. Again, the differences in the
low-energy mode may be attributed to the defect molecule
approximation. Regarding the isotope shifts, the three defects
(C$_{2}$)$_{\text{Hex}}$ in 3C-SiC and
(C$_{\text{sp}}$)$_{2,\text{hk,cub}}$ and
(C$_{\text{sp}}$)$_{2,\text{hh}}$ in 4H-SiC also reveal their common
nature. As expected for a stretching vibration of a carbon-carbon
dumbbell, the highest mode shows a threefold isotope splitting. We
obtain frequency ratios for all three defects of 1.016 for the medium
$^{12}$C$-^{13}$C line and 1.035 [(C$_{\text{sp}}$)$_{2,\text{hh}}$:
1.034] for the $^{13}$C$-^{13}$C line. The energetically lower modes
show a complex broadening without a clear structure.

A different spectrum is obtained for the linear di-interstitial
configurations, as displayed by (C$_{\text{sp}}$)$_{2,\text{kk,lin}}$.
The five LVMs reach up to 203.2\,meV, which is attributed to the short
bond length between the atoms of the linear carbon chain. The
vibrational pattern of this defect is difficult to resolve. In
general, the different modes of this defect are stretching vibrations
of two neighboring carbon atoms against each other. The two outmost
pairs are associated with the high-energy vibrations. The complex
vibrational pattern leads to complex isotope shifts: the highest mode
at 203.2\,meV shows a sixfold and the second highest mode a fourfold
splitting. The two lower modes broaden by up to 6\,meV without
structure.

\subsection{Carbon tri- and tetra-interstitials}
\label{sec:carbon-tri-tetra}

The di-interstitials can trap further carbon interstitials and form
larger carbon aggregates. We investigated the growth of small carbon
clusters of up to four carbon interstitials. We found that in 3C-SiC
the pair of carbon split-interstitials (C$_{\text{sp}}$)$_{2}$ binds
additional carbon atoms most effectively. The reason is that the
interstitial pair (C$_{\text{sp}}$)$_{2}$ possesses dangling bonds
located at the two $sp^{2}$-hybridized carbon atoms which point
towards the adjacent carbon sites (cf. Fig.~\ref{fig:3C-int}a center).
The dissociation energy of the resulting tri-interstitial
(C$_{\text{sp}}$)$_{3}$ amounts to 3.0\,eV (cf.
Tab.~\ref{tab:cluster-bind}). This defect exhibits a negative-$U$
effect with an ionization level $\epsilon(2^{+}|0) = 0.74$\,eV and is
negatively charged for $\mu_{\text{F}} > 1.5$\,eV.  The
tri-interstitial may capture a further carbon interstitial and form a
very stable tetra-interstitial (C$_{\text{sp}}$)$_{4}$, whose
dissociation energy amounts to 5.7\,eV (cf. Fig.~\ref{fig:3C-int}b
left and Tab.~\ref{tab:cluster-bind}). The high stability of
(C$_{\text{sp}}$)$_{4}$ originates from the complete
$sp^{3}$-hybridization of all carbon interstitials, which also renders
this defect electrically inactive.

The hexagonal di-interstitial (C$_{2}$)$_{\text{Hex}}$ in 3C-SiC is
more stable than (C$_{\text{sp}}$)$_{2}$. It can also trap additional
carbon atoms, but with a smaller binding energy. An extra carbon atom
at a neighboring carbon site, forming the tri-interstitial
(C$_{2}$)$_{\text{Hex}}-$C$_{\text{sp}}$, needs only 0.4\,eV to
separate. As a consequence, we expect the kinetic formation of
tri-interstitials in the form of (C$_{\text{sp}}$)$_{3}$, despite the
higher stability of (C$_{2}$)$_{\text{Hex}}$ compared to
(C$_{\text{sp}}$)$_{2}$. Since the reorientation from
(C$_{\text{sp}}$)$_{2}$ towards (C$_{2}$)$_{\text{Hex}}$ involves the
breaking of two carbon-silicon bonds, the capture of a third carbon
interstitial may be possible before the reorientation occurs. The
reorientation of (C$_{\text{sp}}$)$_{2}$ into (C$_{2}$)$_{\text{Hex}}$
and the capture of a third carbon interstitial are hence competing
processes.

Once the tri-interstitial (C$_{2}$)$_{\text{Hex}}-$C$_{\text{sp}}$ has
formed, adding a further carbon atom stabilizes the structure: the
resulting tetra-interstitial ((C$_{2}$)$_{\text{Hex}}$)$_{2}$ has a
dissociation energy of 4.0\,eV (cf. Fig.~\ref{fig:3C-int}b right and
Tab.~\ref{tab:cluster-bind}). It consists of two neighboring hexagonal
di-interstitials forming a rectangle with side lengths of 1.45\,\AA\ 
(di-interstitial within the hexagon) and 1.55\,\AA. The narrow sides
of the rectangle form isosceles triangles with the neighboring carbon
atom from the enclosing hexagon with a side length of 1.43\,\AA.

\begin{figure}

  \includegraphics[angle=270,width=0.7\linewidth]{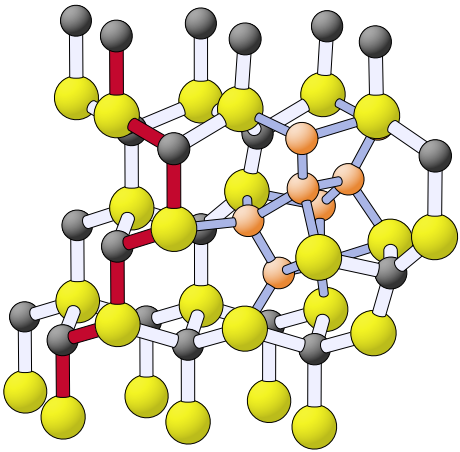}
  \caption{\label{fig:4H-tetra-int} Tetra-interstitial in 4H-SiC
    spanning a cubic and a hexagonal plane. The stacking sequence of
    the crystal is highlighted.}
\end{figure}

Similar interstitial clusters are also available in 4H-SiC. Due to the
enormous amount of possible configurations one has to restrict the
analysis to some sample processes. Similar to 3C-SiC, the aggregation
at a hexagonal di-interstitial such as (C$_{2}$)$_{\text{Hex}}$
results in the lowest formation energy. A split-interstitial bound to
(C$_{\text{sp}}$)$_{2,\text{hh}}$, which resembles---as mentioned
above---a hexagonal di-interstitial, needs only 0.6\,eV to dissociate.
The di-interstitial (C$_{\text{sp}}$)$_{2,\text{kh}}$ in a cubic plane
traps carbon interstitials more effectively: the energy gain for a
split-interstitial at a neighboring hexagonal site amounts to 1.3\,eV
(cf. Tab.~\ref{tab:cluster-bind}). As in 3C-SiC the most efficient
seed for the carbon aggregation is (C$_{\text{sp}}$)$_{2,\text{kk}}$.
The dissociation energy of (C$_{\text{sp}}$)$_{3,\text{kkh}}$ located
in the hexagonal plane amounts to 2.9\,eV (cf.
Tab.~\ref{tab:cluster-bind})---a value comparable to the
split-interstitial cluster (C$_{\text{sp}}$)$_{3}$ in 3C-SiC. With a
fourth carbon interstitial another stable extended defect cluster is
formed. The structure, spanning the cubic and the hexagonal plane, is
shown in Fig.~\ref{fig:4H-tetra-int}. The energy of 5.3\,eV (4.8\,eV)
is needed to remove a carbon atom at a hexagonal (cubic) site. This
defect has a different geometrical structure than
(C$_{\text{sp}}$)$_{4}$ in 3C-SiC. However, all atoms are also
$sp^{3}$ hybridized, which again leads to the high stability of the
complex. Further defect structures can exist in 4H-SiC due to the
alternating cubic and hexagonal planes.  We expect similar
dissociation energies for all these defects.

We analyzed the LVMs of the tri-interstitial (C$_{\text{sp}}$)$_{3}$
and the tetra-interstitials (C$_{\text{sp}}$)$_{4}$ and
((C$_{2}$)$_{\text{Hex}}$)$_{2}$ in 3C-SiC. All defects show
vibrational modes above the SiC bulk phonon spectrum, but for
(C$_{\text{sp}}$)$_{3}$ and (C$_{\text{sp}}$)$_{4}$ the frequency of
the highest mode is decreasing with the number of atoms. The reason is
that the bond length between split-interstitials increases and the
bonding is weakening. Starting with the di-interstitial, the highest
LVM drops from 185\,meV to 163\,meV for (C$_{\text{sp}}$)$_{3}$ and
142\,meV for (C$_{\text{sp}}$)$_{4}$ (cf.
Tab.~\ref{tab:int-clust-lvm}). The highest modes 3 and 4 for
(C$_{\text{sp}}$)$_{3}$ are a anti-symmetric and symmetric stretching
vibrations of the two split-interstitials located at the border of the
triangle-shaped defect complex. Modes 2 and 3 are also antisymmetric
and symmetric vibrations of the pairs which form the sides of the
triangle, whereas in mode 1 the center split-interstitial vibrates
against its outer neighbors. The highly symmetric structure of
(C$_{\text{sp}}$)$_{4}$ is also reflected in the LVMs: the highest
mode 5 describes a stretching vibration of all split-interstitials
where two neighboring interstitials are always in opposite phase. The
modes 3 and 4 are degenerate and represent an antisymmetric vibration
of two opposite split-interstitials, respectively. In mode 2 the long
carbon-carbon bonds between two neighboring split-interstitials are
stretched and in mode 1 the diagonally opposite split-interstitials
rotate against each other.

The complex ((C$_{2}$)$_{\text{Hex}}$)$_{2}$ shows a rich spectrum of
eight LVMs up to 196\,meV. The reason for these high-frequency LVMs is
the compact size of the defect. The two highest modes 7 and 8 are
symmetric and antisymmetric stretching vibrations of the two
triangles. The modes 5 and 6 are a symmetric and antisymmetric tilting
of the narrow sides of the rectangle against the third carbon atoms in
the triangles. Mode 4 describes a breathing vibration of the rectangle
combined with an oscillation of the carbon atoms at the triangles'
apex against each other, whereas mode 3 is an antisymmetric stretching
and quenching of the two triangles. The low-frequency modes 1 and 2
are complex oscillations, which presumably extend beyond the analyzed
defect molecule.

\section{Antisite clusters}
\label{sec:antisite-clusters}

\begin{figure*}
  \includegraphics[width=\linewidth]{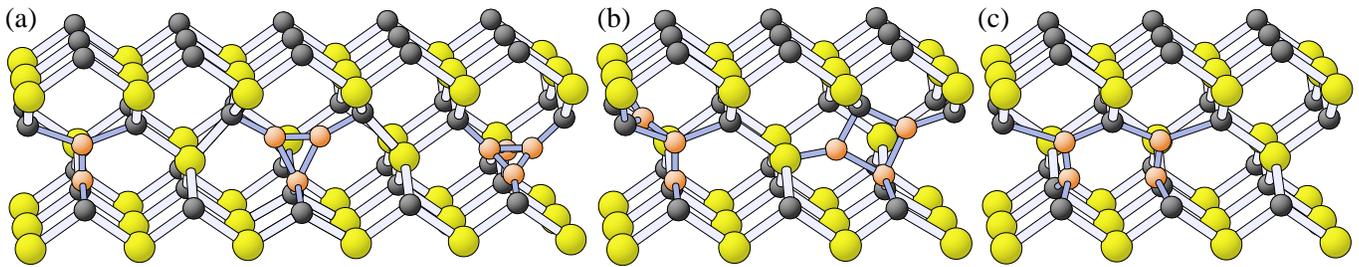}
\caption{\label{fig:antisite-clust} Aggregation of carbon clusters at
  a carbon antisite in 3C-SiC. (a) Dicarbon antisite
  (C$_{2}$)$_{\text{Si}}$, tricarbon antisite (C$_{3}$)$_{\text{Si}}$
  and tetracarbon antisite (C$_{4}$)$_{\text{Si}}$. (b) Intermediate
  configurations $I1$ and $I2$ of the tricarbon antisite. (c) Pairing
  of two dicarbon antisites ((C$_{2}$)$_{\text{Si}}$)$_{2}$.}
\end{figure*}

The carbon antisite has the formation energy of 4.2\,eV (Si-rich
conditions) in 3C-SiC as well as in 4H-SiC. Alongside with the carbon
vacancy, it is the energetically most favorable intrinsic defect.
Various antisite formation mechanisms have been suggested sofar: the
recombination of the $V_{\text{C}}$-C$_{\text{Si}}$ complex with a
silicon atom,\cite{Ma01b} a kick-out mechanism,\cite{Eb02} the
recombination of a carbon interstitial with a silicon vacancy and the
incorporation in as-grown material are likely. Unfortunately, this
defect is hardly detectable: neither does it possess defect states in
the band gap, nor does it show localized vibrational
modes.\cite{Ma01b} Nevertheless, the carbon antisite is of a
particular importance as a nucleus for larger carbon aggregates. It
may capture carbon interstitials and form small, tightly bound
complexes that may be featured as various thermally stable defect
centers.\cite{Ma01b,Ga02,Ga03a,Ma03}

\subsection{Dicarbon antisite}
\label{sec:di-carbon-antisite}

\begin{figure}
  \includegraphics*[width=\linewidth]{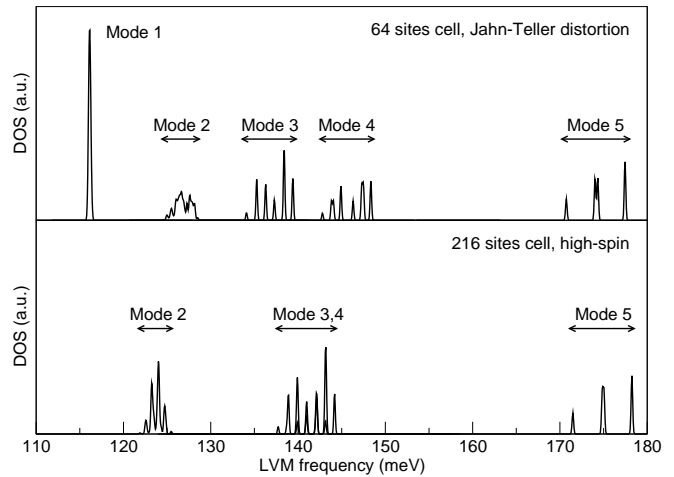}
  \caption{\label{fig:ICSpSi-isospectrum} Isotope shifts of the
    dicarbon antisite enriched with 30\% $^{13}$C. The upper panel
    shows the splitting of the modes calculated in a 64 sites
    supercell in the Jahn-Teller distorted configuration, the lower in
    216 sites supercell with a high-spin like symmetric
    configuration. The frequencies of the pure $^{12}$C case are given
    in Tab.~\ref{tab:anti-clust-lvm}.}
\end{figure}

The most simple and most discussed defect of this kind is the
di-carbon antisite
(C$_{2}$)$_{\text{Si}}$,\cite{Ma01b,Ga02,Ma02,Ma03,Ga03a} where two
carbon atoms share a single silicon site (cf.
Fig.~\ref{fig:antisite-clust} left). It exists in 3C- and 4H-SiC and
has a dissociation energy of 4.1\,eV in 3C-SiC and 3.5\,eV (3.6\,eV)
in 4H-SiC at the cubic (hexagonal) site (cf.
Tab.~\ref{tab:cluster-bind}). The defect is neutral in the Fermi-level
range of $0.76 < \mu_{\text{F}} < 1.11$ in 3C-SiC and $1.27 <
\mu_{\text{F}}< 2.04$ in 4H-SiC. It can prevail in a high-spin state
and a Jahn-Teller distorted low-spin state. The high-spin state is
favored by 80$-$120\,meV, whereas this energy difference is too low to
uniquely determine the ground state. We have discussed the structural,
electronic and vibronic properties of this defect in detail in
Refs.~\onlinecite{Ma01b,Ma02,Ma03}. Here we present the results for
the LVM's isotope shifts of this defect. The LVMs of
(C$_{2}$)$_{\text{Si}}$ for 3C-SiC and 4H-SiC are listed in
Tab.~\ref{tab:anti-clust-lvm}. For the former the results of the
Jahn-Teller distorted configuration in a 64 sites cell and the results
for a high-spin nearly symmetric configuration in a 216 sites cell are
given. The results for 4H-SiC are for the low-spin state and have been
calculated in a 128 sites cell using the defect molecule approximation
(the results for the high-spin state are given in
Ref.~\onlinecite{Ma03}). In Figure~\ref{fig:ICSpSi-isospectrum} we
show the isotope effect for LVMs of (C$_{2}$)$_{\text{Si}}$ in 3C-SiC
assuming a $^{13}$C enrichment of 30\%. We included the dumbbell and
its four nearest neighbors in the calculation and assumed that
$^{13}$C can substitute with equal probability of 30\% any of these
sites. Note that the height of the peaks in
Fig.~\ref{fig:ICSpSi-isospectrum} does not necessarily reflect the
line intensity in the photo-luminescence experiments. We find that
similar to the carbon split-interstitial the energetically highest
mode 5 splits into three lines. The reason is the strong localization
of the vibration on the carbon dumbbell. The middle shifted line, with
a ratio of 1.019 to the highest line in isotope-undoped sample,
results from the vibrations of the $^{12}$C$-^{13}$C dumbbell, whereas
the lowest line, shifted by 1.040, comes from the $^{13}$C$-^{13}$C
dumbbell. This result is identical for both high-spin and Jahn-Teller
distorted configurations. The modes 3 and 4, which are vibrations of
the upper and lower part of the dumbbell against the neighbors, split
into six nearly equidistant lines with a separations between
1$-$1.4\,meV. In the Jahn-Teller distorted case (cf. upper panel of
Fig.~\ref{fig:ICSpSi-isospectrum}) the two shifted spectra are clearly
separated, in the symmetric high-spin case (cf.  lower panel of
Fig.~\ref{fig:ICSpSi-isospectrum}) they overlap and show a distinct
signature of six lines. The mode 2 shows a broadening with two maxima
at 127.6\,meV and 126.7\,meV in the distorted case and four peaks in
the symmetric configuration. In the former, the whole width of the
broadening reaches from 128.5\,meV to 124.9\,meV. The ratios of the
two maxima compared to the $^{12}$C value are 1.007 and 1.014. The
four modes of the latter reach from 124.8\,meV to 122.6\,meV. Mode 1,
which is the breathing mode in Jahn-Teller configuration, remains
practically undisturbed by the $^{13}$C enrichment. This mode,
however, drops into the bulk continuum for the high-spin
case.\cite{Ma03}

\subsection{Clusters of di-carbon antisites}
\label{sec:clusters-di-carbon}

\begin{figure}
  \includegraphics[angle=270,width=0.7\linewidth]{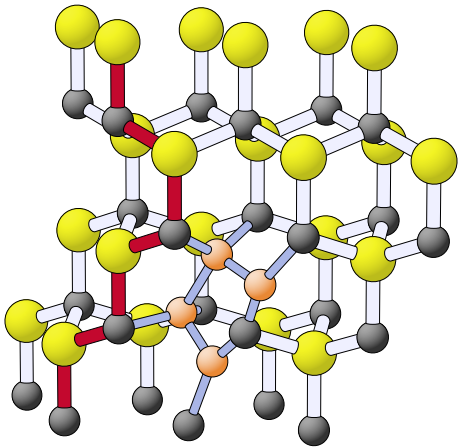}
  \caption{Pair of dicarbon antisites
    ((C$_{2}$)$_{\text{Si}}$)$_{2,kh}$ in 4H-SiC spanning cubic and
    hexagonal sites in the cubic plane. The stacking sequence of the
    crystal is highlighted.}
  \label{fig:cspsi2-4H}
\end{figure}

Another defect, which is closely related to the di-carbon antisite, is
the di-carbon antisite pair ((C$_{2}$)$_{\text{Si}}$)$_{2}$ (cf.
Fig.~\ref{fig:antisite-clust}c). Like the di-carbon antisite, this
defect has already been discussed in earlier
publications.\cite{Ma03,Ga03a} Here we want to compare our results for
the dicarbon antisite pair with the structurally very similar
di-interstitials. For the formation of dicarbon antisite pairs, the
lone pairs or other clusters of antisites have to be created first.
Eberlein \emph{et al.}\cite{Eb02} suggested the antisite formation via
kick-out reactions. Rauls \emph{et al.}\cite{Ra03} discussed a
vacancy-mediated mechanism. The structural difference of the dicarbon
antisite pair from the carbon di-interstitial is in the
carbon-dominated environment. The same defect configurations are
available for the dicarbon antisite pair as for the di-interstitial.
The defect spans two neighboring sites of the same sublattice, hence
in 4H-SiC the same variety of different configurations can form:
cubic-cubic, hexagonal-hexagonal in two different orientations and
cubic-hexagonal within the cubic and the hexagonal plane. Also a
variant comparable to (C$_{2}$)$_{\text{Hex}}$ is available which we
label ((C$_{2}$)$_{\text{Si}}$)$_{2,\text{Hex}}$. It can be visualized
as a hexagonal di-interstitial (Fig.~\ref{fig:3C-int}a right) with a
carbon environment. The di-carbon antisite pair is very stable: the
energy of 5.9\,eV is needed to remove a carbon atom in the 3C-SiC
polytype. This dissociation energy is identical for
((C$_{2}$)$_{\text{Si}}$)$_{2}$ and the hexagonal dicarbon antisite
((C$_{2}$)$_{\text{Si}}$)$_{2,\text{Hex}}$, thus neither configuration
is energetically preferred. In 4H-SiC, the dissociation energy of as
much as 6.7\,eV is obtained for the cubic-hexagonal configuration in
the cubic plane (Fig.~\ref{fig:cspsi2-4H}). Alongside with the
hexagonal-hexagonal configuration, this is the most stable defect.
The calculation of the dissociation energy for all types of the defect
would be too costly. Thus we evaluated the formation energies of the
defects. The formation energy of the cubic-cubic complex is about
0.3\,eV higher than the hexagonal-hexagonal configuration, followed by
the linear structures as described for the di-interstitial (hh:
0.7\,eV, kk: 1\,eV) and the cubic-hexagonal defect in the hexagonal
plane (1.2\,eV). The higher binding energy in 4H-SiC compared to
3C-SiC originates from the enhanced relaxation. The defects thus
benefit from the larger room for the relaxation due to the alternating
cubic and hexagonal planes. Apart from the formation energy, the
defect concentration also depends on the availability of adjacent
carbon antisites and can be kinetically limited.

The LVMs of ((C$_{2}$)$_{\text{Si}}$)$_{2}$ in 3C and 4H-SiC are given
in Tab.~\ref{tab:anti-clust-lvm}. A detailed description of the modes
is presented in Ref.~\onlinecite{Ma03}. Remarkable is the
polytype-independent small splitting of the modes 5 and 6, which is
observed for all configurations of the dicarbon antisite pair. Similar
results are obtained for the cubic-cubic configuration, which is not
listed in Tab.~\ref{tab:anti-clust-lvm}. For this defect, the LVMs
differ by at most 2\,meV from the values of the hexagonal-hexagonal
configuration. The isotope splitting of all configurations is complex.
The defect does not possess isolated dumbbell oscillations, and the
LVMs 5 and 6 are very close. Consequently, the isotope shifts of the
two highest LVMs overlap, resulting in broad peaks with a width of up
to 2.5\,meV which cannot be clearly separated.

The hexagonal dicarbon antisite pair
((C$_{2}$)$_{\text{Si}}$)$_{2,\text{Hex}}$ shows vibrational
properties similar to the hexagonal di-interstitial. The highest LVM
at 174.2\,meV is practically an isolated dumbbell vibration, whereas
the small splitting between the modes 5 and 6 is typical for
((C$_{2}$)$_{\text{Si}}$)$_{2}$. The mode 5 is a rotation of the
dumbbell in the plane of the nearest neighbors. The lower modes are
complex vibrations of the entire carbon surrounding. The complex
vibrational pattern is reflected in the isotope splittings. The mode 6
shows a threefold carbon dumbbell splitting, where the medium mode is
again split into two separate lines with 0.5\,meV energy difference.
The frequency ratios are 1.013 (averaged over the two lines---1.011
and 1.015 otherwise) and 1.031.  This is, however, superimposed with
the complex splitting pattern of the modes 4 and 5. Thus an extensive
isotope broadening between 160\,meV and 175\,meV is expected for this
defect.

Larger aggregations of dicarbon antisites are also conceivable.
Similar to the tetra-carbon interstitial cluster, the complex of
dicarbon antisites can be imagined. Provided the lone antisites are
formed kinetically, the antisite cluster is strongly bound: the energy
of 7.4\,eV is needed to remove a carbon atom from
((C$_{2}$)$_{\text{Si}}$)$_{4}$. The defect is electrically inactive,
similar to its interstitial counterpart. It possesses 11 LVMs above
the SiC bulk phonon spectrum. Similarly to the tetra-interstitial
(C$_{\text{sp}}$)$_{4}$ four modes lie between 138.5\,meV and
142.9\,meV (cf.  Tabs.~\ref{tab:anti-clust-lvm}
and~\ref{tab:int-clust-lvm}). The additional modes in the
low-frequency range up to 124.5\,meV originate from the stronger
carbon-carbon bonds. The bonds are, however, elongated compared to the
dicarbon antisite pair due to the practically undistorted original SiC
surrounding.

\begin{table*}
  \caption{\label{tab:anti-clust-lvm} LVMs of carbon antisite
    clusters in 3C and 4H in meV. The subscripts $k$ and $h$ denote
    the cubic and the hexagonal site. The second values for
    (C$_{2}$)$_{\text{Si}}$ (3C) are from the calculation of
    a full 216 site cell in a symmetric high-spin configuration. The
    results for (C$_{2}$)$_{\text{Si,h}}$ in 4H-SiC are similar to
    (C$_{2}$)$_{\text{Si,k}}$ and given in Ref.~\onlinecite{Ma03}. The
    values marked by an asterisk are slightly below the bulk phonon
    limit, which may originate from the employed defect-molecule
    approximation. The bulk phonon limit obtained from \emph{ab
    initio} calculations without macroscopic crystal polarization is
    115.5\,meV for 3C-SiC and 118.6\,meV for 4H-SiC.}
  \begin{ruledtabular}
    \begin{tabular}{cccccccccccc}
      & \multicolumn{6}{c}{3C} & \multicolumn{4}{c}{4H}\\
      \cline{2-7}\cline{8-12}
      LVM & (C$_{2}$)$_{\text{Si}}$ & (C$_{3}$)$_{\text{Si}}$ &
      (C$_{4}$)$_{\text{Si}}$ & ((C$_{2}$)$_{\text{Si}}$)$_{2}$ &
      ((C$_{2}$)$_{\text{Si}}$)$_{2,\text{Hex}}$ &
      ((C$_{2}$)$_{\text{Si}}$)$_{4}$ & (C$_{2}$)$_{\text{Si,k}}$ &
      (C$_{3}$)$_{\text{Si,k}}$ & (C$_{4}$)$_{\text{Si,k}}$ &
      ((C$_{2}$)$_{\text{Si}}$)$_{2,\text{kh}}$ &
      ((C$_{2}$)$_{\text{Si}}$)$_{2,\text{hh}}$\\
      \hline
      1 & 116.4         & 118.6 & 124.7 & 123.3 & 118.6 & 116.3 & 102.3* & 119.0  & 113.4* & 114.4* & 114.3* \\
      2 & 128.5 / 125.5 & 129.8 & 130.6 & 125.6 & 134.0 & 116.4 & 119.7  & 130.2  & 114.3* & 119.6  & 120.4 \\
      3 & 139.4 / 143.2 & 148.7 & 138.0 & 127.1 & 138.3 & 117.6 & 135.0  & 154.0  & 114.5* & 132.2  & 130.8 \\
      4 & 148.3 / 144.2 & 181.3 & 165.8 & 139.3 & 164.1 & 119.7 & 139.1  & 182.3  & 120.3  & 147.0  & 145.7 \\
      5 & 177.5 / 178.3 & 248.5 & 181.2 & 168.6 & 170.1 & 121.5 & 178.0  & 254.9  & 200.1  & 160.2  & 159.7 \\
      6 &               &       & 189.4 & 169.7 & 174.2 & 121.6 &        &        & 201.5  & 162.4  & 161.8 \\
      7 &               &       & 223.6 &       &       & 124.1 &        &        & 204.3  &&\\
      8 &               &       &       &       &       & 138.5 &        &        & 240.7  &&\\
      9 &               &       &       &       &       & 141.0 &        &&&&\\
     10 &               &       &       &       &       & 142.2 &        &&&&\\
     11 &               &       &       &       &       & 142.9 &        &&&&\\
    \end{tabular}
  \end{ruledtabular}
\end{table*}

\subsection{Tri- and tetra-carbon antisites}
\label{sec:tri-tetra-carbon}

In addition to the clustering, the dicarbon antisites can trap carbon
interstitials. Examples are the tricarbon and the tetracarbon antisite
(cf. Fig.~\ref{fig:antisite-clust}a center and right). Both are
tightly bound defects with three or four carbon atoms at a single
silicon site.

The tricarbon antisite is a triangle-shaped defect with its base
oriented in $\langle 110 \rangle$ direction. The capture of a carbon
interstitial by the dicarbon antisite yields the energy of 4.8\,eV
(cf. Tab.~\ref{tab:cluster-bind}). This capture may occur via several
intermediate configurations, which are depicted in
Fig.~\ref{fig:antisite-clust}b.  First, a carbon interstitial is
trapped at a neighboring carbon site (Fig.~\ref{fig:antisite-clust}b
left, configuration $I1$) with an energy of 1.4\,eV. This is about
1\,eV higher than the migration barrier for the neutral carbon
split-interstitial.\cite{Bo03b} In a second step, a square
configuration of the tricarbon antisite may form (cf.
Fig.~\ref{fig:antisite-clust}b right, configuration $I2$), with the
additional energy gain of 2.2\,eV ($E_{\text{D}} = 3.6$\,eV). This
configuration is metastable and has also been found by Gali \emph{et
  al.}\cite{Ga03a} with a dissociation energy of 4.0\,eV. In a last
step, the triangular configuration is finally formed (configuration
(C$_{3,\text{tri}}$)$_{\text{Si}}$), yielding an additional 1.2\,eV.
This process is also possible in 4H-SiC---an adjacent
split-interstitial to the dicarbon antisite is bound with 1.9\,eV
(configuration $I1$ in Tab.~\ref{tab:cluster-bind}), yielding an
additional energy gain of 3.8\,eV in the triangular configuration
($E_{\text{D}}=5.8$\,eV, configuration
(C$_{3,\text{tri}}$)$_{\text{Si}}$)). In its first formation step the
tricarbon antisite is electrically neutral for $\mu_{\text{F}} >
0.7$\,eV. This value reduces in the next steps.

The defect shows a pronounced LVM spectrum (cf.
Tab.~\ref{tab:anti-clust-lvm}): the five modes reach up to 248.5\,meV
(255.1\,meV in 4H-SiC). This is the highest value among all defects we
have analyzed. This is also much higher than the LVMs in the
intermediate configuration (Fig.~\ref{fig:antisite-clust}b right)
obtained by Gali \emph{et al.}\cite{Ga03a} at 190, 172, 166, 135, 133,
and 117\,meV. The reason for the high frequency is the short bond
length between the adjacent carbon atoms---it ranges from 1.29\,\AA\ 
for the base of the triangle to 1.55\,\AA\ for the distance between
the triangle's apex and the carbon atoms of the enclosing tetrahedron.
The highest mode (mode 5) is a stretching vibration of the short base
of the triangle. Mode 4, which is with 181.3\,meV still very high, is
an asymmetric stretching vibration of the two base atoms against their
nearest neighbors from the enclosing tetrahedron. In mode 3 the
triangle is stretched, whereas in mode 2 the whole triangle oscillates
against the tetrahedron. Mode 1 is a vibration of the carbon atom at
the apex against its neighbors.

Since the mode 5 is strongly localized at the base of the triangle,
one expects a very distinct pattern of isotope shifts similar to the
isolated carbon dumbbell. Indeed, replacing the nearest neighbors of
the triangle with $^{13}$C does not affect this mode at all. Thus the
analysis of the isotope shifts for this mode can be restricted to the
defect itself. Three frequencies are obtained: 248.5\,meV for the pure
$^{12}$C case (248.3\,meV if the carbon atom at the apex is
substituted), 244.3\,meV if one of the two carbon atoms of the base is
replaced by $^{13}$C and 239.8\,meV in the $^{13}$C$-^{13}$C case.
The corresponding ratios are thus 1.018 and 1.036, respectively.  In
4H-SiC, we find the shifted lines at 250.5\,meV and 246.0\,meV, with
frequency ratios of 1.017 and 1.036.  As discussed above, the
energetically lower modes have complex vibrational patterns involving
also the nearest neighbor atoms of the defect. As seen for the
dicarbon antisite, only a broadening of the modes and not such a
distinct splitting will result from substituting the defect's atoms
with $^{13}$C.

With the capture of another carbon interstitial the tetracarbon
antisite (C$_{4}$)$_{\text{Si}}$ is formed (cf.
Fig.~\ref{fig:antisite-clust}a right). With the dissociation energy of
2.8\,eV in 3C-SiC and 2.9\,eV in 4H-SiC (cf.
Tab.~\ref{tab:cluster-bind}) this defect is less stable than
(C$_{3}$)$_{\text{Si}}$, but still tightly bound. In 3C-SiC, the
central carbon dumbbell is oriented in \{110\}-direction having two
additional carbon atoms at each side. The dumbbell is 1.70\AA\ long,
the sides of the two isosceles triangles formed by the dumbbell and
the additional carbon atoms are 1.38\,\AA\ and 1.44\,\AA,
respectively.  For a Fermi-level $\mu_{\text{F}} > 0.6$\,eV and
$\mu_{\text{F}} < 1.5$\,eV the defect is neutral. The asymmetric
geometry is the result of a Jahn-Teller effect. In the symmetric
configuration, the defect possesses a doubly degenerate localized
state in the band gap. Since in the neutral state this level is
occupied, the symmetry is broken by a Jahn-Teller distortion, which
yields an energy gain of about 1\,eV. The asymmetric configuration
gives rise to a rich LVM spectrum with high-frequency LVMs. We have
found 7 LVMs above the bulk spectrum with the highest LVM at
223.6\,meV. The mode 7 is a stretching vibration of the smaller of the
two triangles. The atom at the apex oscillates against the central
dumbbell and the carbon atom of the enclosing tetrahedron. The mode 6
is an antisymmetric stretching vibration of the two atoms of the
central dumbbell atoms against their neighbors. Since the vibration is
antisymmetric, the length of the central dumbbell remains constant.
The mode 5 is a stretching vibration of the second triangle with the
larger sides, whereas the mode 4 is mainly a stretching vibration of
the lateral carbon farther away from the central dumbbell against its
neighbor along the $z$-axis. The modes 1 to 3 describe a complex
motion of the inner carbon atoms against the enclosing tetrahedron.
We find an at least sixfold isotope splitting of the highest mode and
a structureless broadening of the lower modes.

The tetracarbon antisite in 4H-SiC shows a simpler LVM spectrum with
only five LVMs above the bulk phonon spectrum. This is due to the
highly symmetric defect geometry, which is, as mentioned above, not
stable in 3C-SiC. In the low-symmetry 4H crystal, the degeneracy of
the defect states is lifted already in the symmetric configuration.
The defect forms a small tetrahedron with a side length of 1.43\,\AA,
which is enclosed in the carbonic tetrahedron of the crystal lattice.
The highest mode 8 is found at 241\,meV, followed by three nearly
degenerate modes between 200\,meV and 205\,meV. Mode 8 is the
breathing mode of the tetrahedron. The modes 5 to 7 are nearly
degenerate and cannot be separated into vibrations of individual
carbon pairs. The low-frequency modes 1 to 4 show complex vibrational
patterns involving all atoms of the defect and the environment. These
modes are the least localized. With the partial substitution by
$^{13}$C, the mode 8 splits into at least four lines separated by
about 2\,meV.  The modes around 200\,meV broaden into a continuum
about 10\,meV wide.

\section{Discussion}
\label{sec:discussion}

The calculated vibrational spectra advise a comparison with
photo-luminescence experiments. Several defect centers, which have
been attributed to intrinsic defects, may be relevant. There are the
alphabet-lines,\cite{EgHeIv99} the D$_{\text{I}}$ center,\cite{Pa72},
the D$_{\text{II}}$ center\cite{Pa73} and the recently discovered
PL-centers P$-$U.\cite{Ev02} However, the alphabet-lines and the
D$_{\text{I}}$-center possess LVMs only in the phonon band gap. Hence
the defects considered above are not suitable candidates. We will thus
focus on the centers with vibrational modes above the SiC bulk phonon
spectrum.

\subsection{The D$_{\text{II}}$-center}
\label{sec:d2-center}

The D$_{\text{II}}$ center shows vibrational modes above the SiC bulk
spectrum up to 164\,meV, which are to a large extent polytype
independent.\cite{Sr98} Five modes have been described as
``universal'' throughout the polytypes. Yet, a comparison of the
spectra for different polytypes indicates that, in fact, a larger
number of modes can be regarded as polytype-independent.\cite{Ma03}
The number of reported LVMs varies. Five to six modes with tiny peaks
in-between have been counted in 3C-SiC and up to 12 modes have been
observed in 4H-SiC.\cite{Ca03} Another important feature is the
thermal stability of the center: it persists temperatures up to
1700$^{\circ}$C.\cite{Pa73,Fr87,Sr98} First we summarize our results
of Ref.~\onlinecite{Ma03} related to the dicarbon antisite
(Sec.~\ref{sec:di-carbon-antisite}) and to the dicarbon antisite pair
(Sec.~\ref{sec:clusters-di-carbon}). In this context, we consider the
new information concerning the larger defect clusters.

The dicarbon antisite possesses an LVM pattern compatible with the
universal spectrum of the D$_{\text{II}}$-center.\cite{Ma01b} The high
dissociation energy suggests a high thermal stability of this defect.
Yet, the large number of LVMs in the hexagonal polytypes cannot be
reproduced. Therefore, a larger complex, i.e.\ the dicarbon antisite
pair, has been proposed by Gali \emph{et al.}\cite{Ga03a} The idea was
that the combination of the cubic and the hexagonal sites could
explain the rich multitude of zero phonon lines (ZPLs) in 4H and
6H-SiC. However, also this model does not explain in full the features
of the D$_{\text{II}}$-center.\cite{Ma03} Especially, it predicts a
small splitting of the two highest LVMs (cf.
Tab.~\ref{tab:anti-clust-lvm}) for all polytypes and all combinations
of the site pairs, which is not observed in the D$_{\text{II}}$
spectrum. In addition, the number of LVMs is still too low to explain
the D$_{\text{II}}$ spectrum in hexagonal polytypes. In an
experimental paper,\cite{Sr01} it was suggested that the multiple ZPLs
stem from the excited defect states. Yet, the dicarbon antisite model
qualitatively reproduces the most robust features of D$_{\text{II}}$.
In general, the carbon antisite related defects possess LVMs in
necessary frequency range. One can thus expect the dicarbon antisite
to be a major building block or precursor of the
D$_{\text{II}}$-center.  However, the dicarbon antisite itself and the
dicarbon antisite pair are too small to explain the whole vibrational
spectrum. A larger defect structure seems indispensable, which could
also account for the polytype-independent features.

On the other hand, the interstitial clusters as the model for the
D$_{\text{II}}$-center can be ruled out. None of the clusters shows a
proper LVM spectrum. For the di-interstitials, a few LVMs above
160\,meV exist, as well as a mode around 130\,meV. The highest LVM has
the tendency to drop with growing cluster size. Therefore, adding
further interstitials would not improve the situation. On the
contrary, the carbon-rich environment due to the incorporation of
antisites provides the diamond-like spectrum of the
D$_{\text{II}}$-center. This is seen for the clusters at a single
antisite and the di-carbon antisite aggregates: the trapping of
further carbon interstitials yields very short carbon-carbon bonds
with high-frequency LVMs above 200\,meV, whereas a clustering of
dicarbon antisites results in a dropping of the highest mode, but also
an increase of the number of LVMs. The combination of both processes
may thus explain the LVM spectrum of the D$_{\text{II}}$-center.

\subsection{The P$-$T centers}
\label{sec:p-t-centers}

Further intrinsic defects with high-frequency LVMs are the five
centers labeled P$-$T, which were found in electron-irradiated
6H-SiC.\cite{Ev02} An electron beam with the energy below the silicon
displacement threshold, but at a very high dose of
$3\times10^{20}\,e$~cm$^{-2}$ was used. The five centers possess
separate ZPLs with LVMs at 133\,meV and 180\,meV. The P-center is the
most stable and persists annealing temperatures of 1000$^{\circ}$C.
Isotope-enriched PL-measurements with a $^{13}$C incorporation of 30\%
were performed. The higher LVM splits into three lines, with the
isotope-undoped line shifted to about 175\,meV and 171\,meV. This
results in frequency ratios (averaged over the five different centers)
of 1.020$\pm 0.002$ and 1.043$\pm 0.002$. This isotope splitting has
been interpreted as a vibration of a carbon-carbon dumbbell. For this
defect the stretching vibration shows a threefold isotope splitting.
The experimental results are in agreement with the square root
dependence on the reduced mass. The ratios of the vibrational
frequencies of a $^{12}$C$-^{13}$C and a $^{13}$C$-^{13}$C dumbbell to
frequency of the the pure $^{12}$C case are 1.020 and 1.041,
respectively. A simple shift to 128\,meV has been observed for the
lower LVM.  This yields a frequency ratio of 1.034$\pm0.001$.  For the
interpretation of this mode the oscillation of the carbon atom against
its silicon neighbors in a Si$-$C$-$Si molecule has been considered,
which gives a frequency ratio of 1.033.  This underlies a carbon
split-interstitial model proposed for the P$-$T centers.

According to our calculations, the split-interstitials
(Sec.~\ref{sec:carb-split-interst}) and the dicarbon antisite
(Sec.~\ref{sec:di-carbon-antisite}) possess vibrational modes close to
the LVMs of the P$-$T centers. The originally suggested
split-interstitial\cite{Ev02} has a single LVM above the SiC bulk
spectrum in the symmetric case (219.2\,meV in the charge state $2^{+}$
(Ref.~\onlinecite{Ma01b}) and 183\,meV in the neutral charge
state\cite{Ga03a} in 3C-SiC, 197\,meV at the hexagonal site in
4H-SiC\cite{Ga03a}). This picture, however, changes for the
energetically more favorable tilted configuration. In this case two
(three) LVMs at the cubic (hexagonal) site in good agreement with the
values of the P$-$T centers are found
(Sec.~\ref{sec:carb-split-interst}). Also the values of the frequency
ratios for the high-frequency mode are in excellent agreement. This
finding supports the split-interstitial model of Evans \emph{et
  al.}\cite{Ev02} Yet, some issues still have to be resolved: First,
the ratios of the isotope shifts for the low-frequency LVMs
(Tab.~\ref{tab:iso-csp}) are smaller than that of the P$-$T centers,
especially in 3C-SiC. Second, the shifted frequencies and the ratios
depend on the site of the substituted carbon atom. Third, two very
close low-frequency modes are obtained at the hexagonal site. All
these facts are not evident from the experimental data. However, the
spectrum is very complex and difficult to resolve,\cite{Ev02} hence it
is not unlikely that further weak peaks close to the reported
vibrational modes may exist. A further difficulty is the high mobility
of the carbon interstitial,\cite{Bo03b} which is not compatible with
the high annealing temperature of $\approx 1000^{\circ}$C of the
centers.  Two explanations are possible: first, the carbon
interstitials are trapped in interstitial clusters,\cite{Bo04} and
re-emitted in the annealing process. If this is the case, it should be
possible to anneal out the P$-$T centers at lower temperatures, but at
longer annealing times. Second, the low-energy irradiation with a very
high flux produces a large amount of carbon interstitials, which may
not completely annihilate with vacancies during the annealing.  The
mobile interstitials migrate away from vacancies and maintain a
metastable high concentration after the cooling. This interpretation
is supported by the finding that the intensity especially of the
P-center drops very slowly outside the irradiated area.

The dicarbon antisite has been discussed as a possible model by Gali
\emph{et al.}\cite{Ga03a} The mode 5 of (C$_{2}$)$_{\text{Si}}$ (cf.
Tab.~\ref{tab:anti-clust-lvm}) has been associated with the
high-frequency LVM of the P$-$T centers. Due to the localized
character of this vibration of the carbon dumbbell, we observe a
threefold isotope splitting, and the corresponding frequency ratios
indeed agree very well with the experimental results.  The modes 3 and
4 have been associated with the low-frequency LVM (the high-spin case
has been considered in Ref.~\onlinecite{Ga03a}, where these two modes
are degenerate). However, this assignment bears some difficulties.
First, the energy difference between the symmetric high-spin case and
the Jahn-Teller distorted low-spin case is too small to uniquely
determine the true ground state configuration (80\,meV in 3C-SiC and
120\,meV in 4H-SiC in favor of the high-spin state).\cite{Ma03}
Second, due to the lower crystal symmetry in 4H-SiC the modes 3 and 4
are non-degenerate even in the high-spin case.\cite{Ma03} With its
hexagonal-cubic-cubic stacking sequence the polytype 6H-SiC has a
similarly low symmetry, thus it is likely that in 6H-SiC the modes 3
and 4 of the defect should also be observed independently. An
experimental verification in PL measurements is possible by imposing a
uniaxial pressure on a non-enriched crystal. The pressure favors the
distorted configuration and thus the splitting of the modes 3 and 4.
Analyzing the isotope shifts of the modes 3 and 4, Gali \emph{et al.}
substituted the two dumbbell atoms with $^{13}$C.\cite{Ga03a} Hereby a
twofold splitting of the degenerate mode (modes 3 and 4) was observed,
in good agreement with experiment. However, the inclusion of the
nearest neighbor atoms in the isotope analysis has a significant
impact on these modes. In the Jahn-Teller distorted configuration, the
two lines split into two groups with six shifted lines in each (cf.
Fig.~\ref{fig:ICSpSi-isospectrum}). In the symmetric high-spin case
the groups overlap and result in seven lines, which cannot be
separated. Thus this model cannot explain the twofold splitting of the
lower mode of the P$-$T center. The mode 2 was declared invisible by
Gali \emph{et al.}\cite{Ga03a} due to the defect's $D_{2d}$ symmetry.
However, in the Jahn-Teller distorted configuration the symmetry
reduces to $C_{2v}$ and in the hexagonal polytypes to $C_{1h}$.
Similar arguments refer to isolated molecules,\cite{Ca98} which is not
relevant in case of the dicarbon antisite. Yet, the dicarbon antisite
could explain the thermal stability of the P$-$T centers of up to
1000$^{\circ}$C without referring to the cluster formation. The
defect, however, needs carbon antisites as a prerequisite for its
formation.  Since the samples were irradiated with low-energy
electrons, a decent concentration of antisites has to be created by
kinetic processes, before the dicarbon antisite can form.

The five different ZPLs may originate from the various locations of
the defect (either the dicarbon antisite or the split-interstitial) at
inequivalent sites in 6H-SiC. Another explanation may be the
excitation of the defect states in the band gap. The experiments have
been performed with excitation energies below the SiC band
gap,\cite{Ev02} hence it is likely that the bound excitons were formed
by the excitation of an occupied defect state and not by trapping a
free exciton. The energy of the ZPL may thus depend on the energetic
position of the defect level in the band gap, which may vary with the
location of the defect at a cubic or a hexagonal site. Also the
generation of excited exciton states should be considered. This may be
clarified with the measurement of relaxation times. The analysis
performed in Ref.~\onlinecite{Ev02} of the centers' spatial
distribution in the irradiated region and its boundary may also give
an indication of the correlations between the various centers.

Both models (dicarbon antisite and carbon split-interstitial) show
properties that allow to consider them as candidates for the P$-$T
centers. However, both have difficulties that have to be resolved.
Among the other defects that we studied, neither the interstitial
clusters nor the antisite-related clusters can explain the P$-$T
centers. Either the LVM pattern is incompatible with experiment or the
isotope splitting. Considering all these results, the
split-interstitial appears to be the simplest and the most likely
interpretation.

\subsection{The U-center}
\label{sec:u-center}

Besides the P$-$T centers Evans \emph{et al.}\cite{Ev02} have found a
center with a high-frequency mode at 246.6\,meV which has been labeled
the U-center. Further modes have not been identified in the spectra
sofar. In $^{13}$C enriched material two shifted modes at 242.1\,meV
and 237.4\,meV are observed, with the frequency ratios of 1.019 and
1.039, respectively. It has been speculated that the threefold
splitting again originates from the carbon-carbon dumbbell vibrations.
However, such a high vibrational frequency had not been observed for a
carbon-based defect before. Also the frequency ratios do not perfectly
fit the square root dependence on the mass ratios. LVMs with such high
energies have been earlier identified as hydrogen or deuterium (C$-$H
and C$-$D) vibrational modes at 369\,meV and 274\,meV,
respectively\cite{Ch72}. Therefore Evans \emph{et al.}\cite{Ev02}
suggested that hydrogen may be involved in this defect center.

Yet, the calculations for the tricarbon antisite
(C$_{3}$)$_{\text{Si}}$ and the tetracarbon antisite
(C$_{4}$)$_{\text{Si}}$ (cf. Sec.~\ref{sec:tri-tetra-carbon}) show
that LVMs up to 250\,meV should be possible for purely carbon-based
defects. For the tricarbon antisite, the highest mode at 248.5\,meV is
a strongly localized stretching vibration that shows a threefold
isotope splitting with frequency ratios (1.018 and 1.036) close to
that of the U-center. Thus the relation of the U-center to the
carbon-carbon dumbbell is plausible. As discussed in
Sec.~\ref{sec:tri-tetra-carbon}, the vibrations of the energetically
lower modes of the tricarbon antisite are very complex and involve
several carbon atoms, resulting in a strong isotope broadening of
these lines.  In the complex spectra presented in
Ref.~\onlinecite{Ev02}, additional low-frequency lines may be hidden.
If the further lower modes of the U-center were found, an
identification of the U-center with the tricarbon antisite would be a
good possibility. The high dissociation energy of 4.8\,eV in 3C-SiC
and 5.8\,eV in 4H-SiC can also explain the thermal stability of the
U-center, which has been observed up to 1300$^{\circ}$C.

Evans \emph{et al.}\cite{Ev02} also discussed the spatial distribution
of the U-centers. Directly after irradiation the center has been
observed, if at all, in the periphery of the irradiated region, and
inside the region only after annealing above 1000$^{\circ}$C. The
annealing behavior is completely different to that of the P$-$T
centers. This finding supports our interpretation of the preceding
section that the split-interstitials are responsible for the P$-$T
centers, since the U-center and the P$-$T centers must have a
different origin. If we identified the P$-$T centers with the dicarbon
antisite and the U-center with the tricarbon antisite, the P$-$T
centers should coexist in the same region, which apparently is not the
case. The split-interstitial and the antisite defects are, however,
fundamentally different defect configurations. The interpretation of
the dicarbon antisite as a building block of the
D$_{\text{II}}$-center (cf. Sec.~\ref{sec:d2-center} and
Ref.~\onlinecite{Ma03}) implies that also the tricarbon antisite
should then be a precursor of the D$_{\text{II}}$-center. If tricarbon
antisite is related to the U-center, this should be observed in the
experimental data. Indeed, the U-center vanishes at 1300$^{\circ}$C,
whereas the D$_{\text{II}}$-center rises above these
temperatures.\cite{Fr87} Careful annealing studies of the intrinsic
PL-centers could help to clarify the formation behavior of the various
defects.

\subsection{Comparison with carbon aggregation in diamond}
\label{sec:defect-complexes}

In diamond, di-, tri- and tetra-interstitials have been considered by
\emph{ab initio} methods. Goss \emph{et al.}\cite{Go01} have found
similar configurations to (C$_{\text{sp}}$)$_{2}$,
(C$_{2}$)$_{\text{Hex}}$, (C$_{\text{sp}}$)$_{3}$ and
(C$_{\text{sp}}$)$_{4}$. The defects possess a very rich vibrational
spectrum above the diamond bulk phonon limit at 164\,meV. The
di-interstitials show six LVMs up to 228\,meV ($I_{2}$ similar to
(C$_{\text{sp}}$)$_{2}$) and seven LVMs up to 243\,meV ($I_{2}$
similar to (C$_{2}$)$_{\text{Hex}}$). The hexagonal di-interstitial
thus shows a much higher frequency than $I_{2}$ based on two
split-interstitials, which is not so clearly visible in 3C-SiC. The
tri-interstitial $I_{3}$ shows 11 LVMs up to 216\,meV and the
tetra-interstitial $I_{4}$ 12 LVMs up to 197\,meV. It is observed that
the frequencies are generally much higher than in SiC due to the
shorter bond lengths of diamond. The overall behavior is very
similar: the highest LVM goes down with increase of the number of
interstitial atoms, a feature that is also observed for the
interstitial clusters in SiC (cf.  Tab.~\ref{tab:int-clust-lvm}). The
high number of LVMs has been obtained for the ring of dicarbon
antisites ((C$_{2}$)$_{\text{Si}}$)$_{4}$, with all dumbbells enclosed
by a pure carbon shell. For this defect 11 LVMs are counted instead of
five for (C$_{\text{sp}}$)$_{4}$, whereas the highest LVM does not
differ considerably compared to (C$_{\text{sp}}$)$_{4}$. Thus for the
defects spanning several lattice sites only the number of LVMs appears
to depend on the number of carbon atoms involved. The highest LVM
frequency is determined by the crystal lattice.

Another similarity of SiC and diamond is the increase of the
dissociation energy per atom with the cluster size. The cluster
dissociation energy is defined in Ref.~\onlinecite{Go01} as
$E_{\text{D}}= n E_{\text{f}}(I_{1}) - E_{\text{f}}(I_{n})$, where $n$
is the cluster size. For the di-interstitials, this definition gives
the same values as listed in Tab.~\ref{tab:cluster-bind}. For the tri-
and tetra-interstitial we obtain $5.8$\,eV and $11.5$\,eV,
respectively. The increase of the dissociation energy can be
attributed to the larger number of $sp^{3}$-hybridized carbon atoms in
the defect.

The comparison with diamond shows that defects with a high vibrational
frequency (tricarbon and tetracarbon antisite) may exist in SiC. Since
they are confined at a single lattice site, they are only weakly
influenced by the surrounding lattice. The short carbon-carbon bonds
give rise to high-frequency LVMs as in diamond.

\section{Summary and Conclusion}
\label{sec:summary-conclusion}

We investigated the energetic, structural and vibrational properties of
small carbon clusters in 3C and 4H-SiC. We considered two different
types of aggregates. One are clusters of carbon interstitials, where
we investigated clusters with up to four interstitials. The other type
is a collection of carbon atoms at carbon antisite positions. Clusters
comprised of up to four carbon atoms at a silicon site were found. Due
to their short carbon-carbon bonds and their light carbon constituents
they reveal vibrational modes well above the SiC bulk phonon spectrum.
We analyzed the vibrational signatures as well as the isotope shifts
of the defects by \emph{ab initio} calculations and compared them with
the photoluminescence centers P$-$T, U and D$_{\text{II}}$. As the
origin of the P$-$T centers the carbon split-interstitial and the
dicarbon antisite were discussed. Both defects possess LVMs at
appropriate frequencies. The isotope shifts of the high-frequency mode
of the P$-$T centers are in good agreement with the results for both
defects. Yet, the two-fold isotope splitting of the low-frequency mode
cannot be explained by the dicarbon antisite, which shows an at least
sixfold splitting for the modes in that frequency range. For the
split-interstitial the isotope splitting of the low-frequency mode is
compatible with experiment. None of the other interstitial and
antisite clusters showed such a clear isotope splitting as observed
for the P$-$T centers.

The highly mobile carbon interstitials can aggregate and form stable
carbon interstitial clusters. The signatures of di-, tri- and
tetra-interstitials were presented. The most stable form of the
di-interstitial is the hexagonal di-interstitial
(C$_{2}$)$_{\text{Hex}}$. This defect possesses an LVM spectrum which
is practically independent of the polytype and in 4H-SiC also
insensitive to variations of the configuration. The highest LVM of
interstitial clusters reduces with the number of split-interstitials
involved---a behavior that is consistent with similar defects in
diamond. Only very compact interstitial clusters as
((C$_{2}$)$_{\text{Hex}}$)$_{2}$ can have high-frequency LVMs. Carbon
aggregates may also affect the annealing processes of intrinsic
defects. They can trap mobile carbon-interstitials and re-emit them
again at higher temperatures and thus provide the constituents for
further stable defect centers. The apparent contradiction in the
identification of the thermally stable P$-$T centers with a highly
mobile split-interstitial may be lifted in this way.

Also the carbon antisite may trap carbon interstitials. This may
result in very compact defect clusters with LVMs up to 255\,meV, which
was obtained for (C$_{3}$)$_{\text{Si}}$ in 4H-SiC. Consequently,
defect centers like the U-center may be entirely carbon-based. Also
complexes of neighboring antisites may form. Examples are the pair of
dicarbon antisites ((C$_{2}$)$_{\text{Si}}$)$_{2}$ and the ring of
dicarbon antisites ((C$_{2}$)$_{\text{Si}}$)$_{4}$. These defects
behave similar to the split-interstitial clusters, yet they show a
richer spectrum. The relevance of these complexes for the
D$_{\text{II}}$ center was discussed. The compact antisite clusters
cause LVMs well above 200\,meV, whereas the complexes spanning several
defect sites have energetically lower, but richer spectra. Thus for
the D$_{\text{II}}$ center an extended defect which grows on the
dicarbon antisite as core structure may be the origin. This implies
that simple defects like the dicarbon antisite should be observed
experimentally as a precursor of the D$_{\text{II}}$ center.

The centers P$-$U and D$_{\text{II}}$ are yet the only experimentally
observed intrinsic defects with LVMs above the SiC bulk phonon
spectrum. We have presented various stable defect configurations with
high-frequency LVMs. This indicates that probably a number of new
defect centers shall be discovered in future experiments.

\begin{acknowledgements}
  We acknowledge fruitful discussions with J. W. Steeds and M.
  Hundhausen. This work has been supported by the Deutsche
  Forschungsgemeinschaft within the SiC Research Group.
\end{acknowledgements}


\end{document}